\documentclass[fleqn,12pt, a4paper]{article}
\usepackage[dvips]{epsfig}
\usepackage{latexsym}

\def\lsim{\raise0.3ex\hbox{$<$\kern-0.75em\raise-1.1ex\hbox{$\sim$}}}
\def\gsim{\raise0.3ex\hbox{$>$\kern-0.75em\raise-1.1ex\hbox{$\sim$}}}
\newcommand{\AmS}{{\protect\the\textfont2
  A\kern-.1667em\lower.5ex\hbox{M}\kern-.125emS}}

\newcommand{\Tr} {\mbox{Tr}}
\newcommand{\beq}{\begin{equation}}
\newcommand{\eeq}{\end{equation}}
\newcommand{\beqa}{\begin{eqnarray}}
\newcommand{\eeqa}{\end{eqnarray}}

\newcommand{\bn}{\beta/\nu}
\newcommand{\gn}{\gamma/\nu}
\newcommand{\n}{1/\nu}
\newcommand{\Ns}{N_{\sigma}}
\newcommand{\Nt}{N_{\tau}}
\newcommand{\NGsm}{Nambu-Goto string model} 

\begin{document}


\title{  A Study of Finite Temperature Gauge Theory in (2+1)
  Dimensions}

\author{ J. Engels,  F. Karsch,   E. Laermann,   C. Legeland,   \\
         M. L\"utgemeier,   B. Petersson,   T. Scheideler   \vspace*{1ex}  \\ 
         Faculty of Physics, University of Bielefeld,   \\
         P.O. Box 100131, 33501 Bielefeld, Germany
} 

\maketitle 
\thispagestyle{empty}

\begin{abstract}

We determine the critical couplings and the critical exponents of the finite
temperature transition in SU(2) and SU(3)
pure gauge theory in (2+1) dimensions. The critical exponents are in agreement
with those predicted for the universality class of the reduced model for the 
Polyakov loop.
We also measure Wilson
loops at $T=0$ on a wide range of $\beta$ values using APE smearing to 
improve the signal. We extract the string tension $\sigma$ from a fit to 
large distances, including a string fluctuation term.
With these two entities we calculate $T_c/\sqrt{\sigma}$, which is in 
astonishing good
agreement with the \NGsm~for SU(3) and only 7\% off for SU(2).

\end{abstract}

\newpage


\section{Introduction}

The determination of the deconfinement temperature $T_c$ is an important task
in SU(N) gauge theories. The \NGsm~\cite{alvarezolesen} predicts a value for $T_c$ in units of the zero
temperature string tension $\sqrt{\sigma}$ depending only on the dimension but
not on the group.
This has been tested for d=4 and good agreement has been found for N=3 and
especially for N=2 \cite{eos}.
Now we verify the prediction for $T_c/\sqrt{\sigma}$ in 3 dimensions again for
the SU(2) and SU(3) case. 
We determine the critical exponents to check the universality class of the
model. These exponents can be compared with those from the reduced model
(RM), which is in the same universality class as the 2d Ising (SU(2)) 
and the 2d three-state-Potts (SU(3)) model.

We measured $\beta_c $ and the critical exponents on various 
asymmetric  $\Ns^2 \times \Nt$ lattices and $\sqrt{\sigma}$ on a 
symmetric $32^3$ lattice using the standard Wilson action.

In (2+1) dimensions $g_3^2$ has a dimension and sets the scale. Then the bare
lattice coupling is given by
\beq 
  g^2_{3B} = g^2_3 + a g^4_3 + \cdots, \;\;\;
  \beta = \frac{2 N_c}{a g^2_{3B}}
  \label{form:beta}
\eeq

\section{The Phase Transition}

The order parameter for deconfinement on infinite volume lattices is the
Polyakov loop,
\beq
  L = \frac{1}{\Ns^2} \sum\limits_{x} \frac{1}{N_c} 
      \Tr \prod\limits_{\tau = 1}^{\Nt} U_{\nu = 0} (\tau, x).
\eeq  
To find the critical coupling, we look at the Polyakov 
loop $\langle |L| \rangle$, the susceptibility $\chi_L$ 
and the Binder cumulant $g_r$
\beq  
  \chi_L = \Ns^2 \; ( \langle L^2 \rangle -  \langle |L| \rangle^2 )\\
  g_r  = \frac{ \langle L^4 \rangle}{ \langle L^2 \rangle^2} - 3.
\eeq
We use the position of the maximum of the  susceptibility to fix a 
pseudo critical point on a given lattice.
We ran simulations 
on a wide range of spatial lattice sizes ranging 
from $\Ns = 16$ to $96$ for $\Nt = 2$, 
     $\Ns = 16$ to $128$ for $\Nt = 4$ and 
     $\Ns = 48$ to $192$ for $\Nt = 6$ 
at couplings close to the expected
infinite spatial volume critical coupling.

\begin{figure}[htb]
  \center
  \epsfig{file=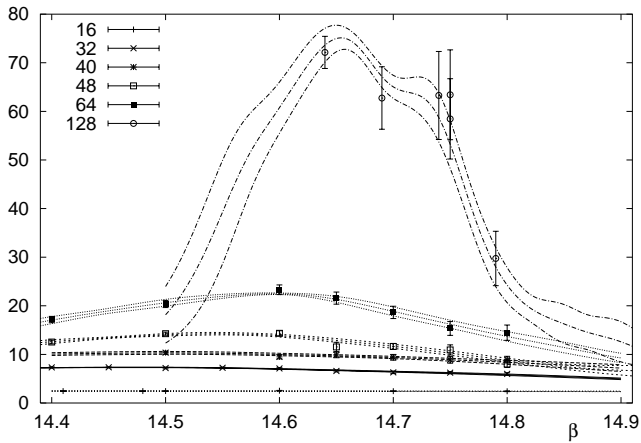, width=90mm }
  \caption{Susceptibility for $\Nt=4$, SU(3) } 
  \label{fig:chi}
\end{figure}

The susceptibility is shown in figure \ref{fig:chi} for SU(3) and $\Nt = 4$. 
The interpolated lines are calculated by the multi histogram density of states
method (DSM). 

To find the critical coupling on an infinite spatial lattice one can
extrapolate with Finite Size Scaling (FSS) ansatz from the values found in
the susceptibility and the Binder cumulant, or use 
the $\chi^2$ Method \cite{Engels}.

The latter uses the FSS formula e.g.~for the susceptibility\footnote{For computational 
  reasons we better take the observable $\chi_L=\Ns^2 \langle L^2 \rangle$. 
See ref. \cite{ExtVersion}} expanded around the critical coupling 
\beq
  \chi_L = \Ns^{\frac{\gamma}{\nu}} 
    ( c_0 +(c_1 +c_2\Ns^{-y_i}) x \Ns^{1/\nu} + c_3\Ns^{-y_i}),
  \label{form:fss}
\eeq
where $x = \frac{\beta_{c, \infty}- \beta}{\beta}$ is the reduced temperature
 that vanishes at the critical coupling,
$c_i$ are coefficients and $y_i$ irrelevant exponents. 
Neglecting corrections $\Ns^{-y_i}$ and taking the logarithm we find 
exactly at $\beta_{c, \infty}$ 
\beq
  \ln{\chi_L} = \ln (c_0) + { \gamma/\nu  } \ln (\Ns).
\eeq
Elsewhere we expect deviations from this as can be seen from
eq.~\ref{form:fss}. 
Thus in a plot of $\ln \chi_L$ vs.~$\ln \Ns$ for fixed $\beta$ the points
should lay on a straight line if (and only if) $\beta=\beta_{c, \infty}$.

Performing a linear fit for every $\beta$ value the one with 
the lowest $\chi^2/dof$ indicates $\beta_{c, \infty}$.

\begin{figure}[htb]
  \center
  \epsfig{file=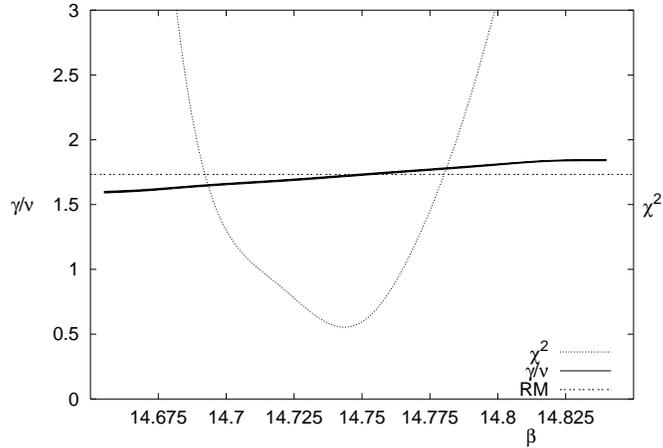, width=90mm }
  \caption{$\beta_{c, \infty}$ and $\gamma/\nu$ from the $\chi^2$ method}
  \label{fig:BetaC}
\end{figure}

In figure \ref{fig:BetaC} the result is shown for the
susceptibility for SU(3), $\Nt = 4$. Plotted is $\gamma/\nu$ vs. $\beta$ (left scale) and $\chi^2$
vs. $\beta$ (right scale). The critical coupling is determined as the $\beta$
value at the minimum of the $\chi^2/dof$ curve. 
An estimate for the error in $\beta_{c, \infty}$ is the min$(\chi^2) +1$, 
this also gives the error on $\beta/\nu$.

\begin{table*}[hbt]
  \setlength{\tabcolsep}{1.5pc} 
  \newlength{\digitwidth} \settowidth{\digitwidth}{\rm 0}
  \catcode`?=\active \def?{\kern\digitwidth}
  \begin{center}
    \begin{tabular}{|l|l|lll|} \hline 
      SU(2)    & $\beta_{c, \infty}$ & $\bn$ & $\gn$ &  $\n$ \\ \hline
      $\Nt =4$ & 6.52 (3) & 0.14 (3) & 1.70 (4) &  0.9 (-1)(+3)    \\
      $\Nt =6$ & 9.55 (4) & 0.17 (3) & 1.68 (7) &  1.3 (4)  \\
      RM       &          & 0.125    & 1.75     &  1.0 \\ \hline \hline
      SU(3)    &  $\beta_{c, \infty}$ & $\bn$ & $\gn$ &  $\n$ \\ \hline
      $\Nt =2$ & ~ 8.155 (15)    & 0.12 (10) & 1.71 (3)   & 1.11 (25)~\\
      $\Nt =4$ & 14.74 (5)     & 0.12 (5)  & 1.72 (7)   & 1.14 (24)~\\
      $\Nt =6$ & 21.34 (4)(11) & 0.14 (2)  & 1.69 (5)   & 1.11 ~\\
      RM       &               & 0.1333    &  1.733     & 1.200 \\ \hline
    \end{tabular}
  \end{center}
  \caption{The critical couplings and critical exponents for SU(2) and SU(3) in
    (2+1) dimensions.}
  \label{tab:CritEx}
\end{table*}

In table \ref{tab:CritEx} the results for the critical couplings and
exponents are summarized.

\section{The String Tension}

We measured Wilson loops on a symmetric $32^3$ lattice.
In SU(2) we simulated only at the critical couplings for $\Nt=4,6$.
In SU(3) we analyzed the potential within a 
wide range of $\beta$ values ranging from 8.15 to 50.
We applied APE smearing \cite{ape} to maximize the signal/noise 
ratio \cite{Smearing}.

With the standard estimator 
\beq
  V_S(R) = \log \frac{\langle {\rm W}(R,S) \rangle}{\langle {\rm W}(R,S+1) \rangle},
\eeq
we get the potential $V(R)$ in the limit of large $S$
\beq
  V(R) = \lim\limits_{T\rightarrow\infty} \; V_S(R).
\eeq

We used the function $ V(R) = V_0 + \frac{\alpha}{R} + \sigma R $ 
for three and two parameter fits.
In the latter case  $\alpha$  was fixed to $\pi/24$ which is the value 
for string fluctuations. This ansatz gave the best and most stable description of
the potentials. 
We performed the fit from an $R_{min}$ to neglect short distance
Coulomb behaviour.

From eq.~\ref{form:beta} we can assume the following scaling behaviour
\beq
  {\rm a} \sqrt{\sigma} 
     =  \frac{c_0}{\beta}+\frac{c_1}{\beta^2}+ \frac{c_2}{\beta^3} + \cdots
  \label{form:fit}
\eeq

\begin{figure}[htb]
  \center  
  \epsfig{file=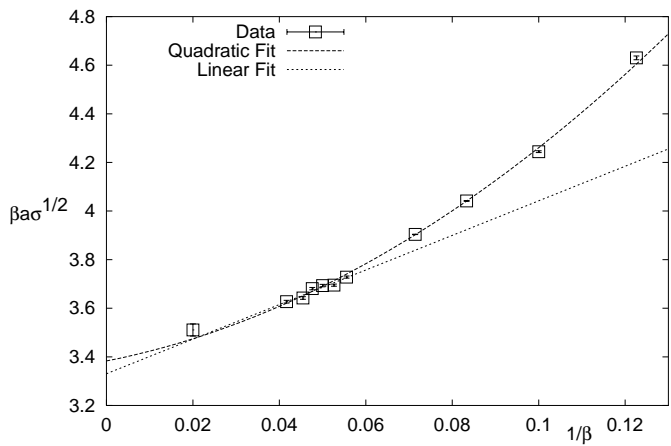, width=90mm }
  \caption{Scaling behaviour of the string tension}
  \label{fig:AsymptoticScaling}
\end{figure}

In figure \ref{fig:AsymptoticScaling} 
we plot $ \beta {\rm a} \sqrt{\sigma}$ versus $1/\beta$. 
The linear fit only contains the six lower data points. 
To describe $ \beta {\rm a} \sqrt{\sigma}$ in the whole range of $\beta$
values one obviously needs a quadratic fit.

\section{$T_c/\sqrt{\sigma}$}

In the case of SU(3) 
we interpolate $\sqrt{\sigma}$ for the 
values $\beta_{c, \infty}$ for the various $\Nt$ from equation \ref{form:fit}. 
For SU(2) we calculate $T_c/\sqrt{\sigma}$ direct from the string tension
obtained at the critical couplings.

Within the \NGsm~one finds
\beq
 T_c/\sqrt{\sigma}  = \sqrt{\frac{3}{\pi(d-2)}} = 
 \cases{ 0.977 & d = 3 \cr
         0.691 & d = 4 }
\eeq 
The results obtained for $T_c/\sqrt{\sigma}$ in (2+1)
dimensions for SU(2) and SU(3) from our lattice calculations are given below
\begin{center}
  \begin{tabular}{lccc}
    &  $\Nt=2$  &  $\Nt=4$   &  $\Nt=6$ \\  
    SU(2)&           & 1.060 (8)  & 1.065 (6) \\
    SU(3)& 0.888(17) & 0.955 (14) & 0.972 (10) \\
  \end{tabular}
\end{center}

\begin{figure}[htb]
  \center
  \epsfig{file=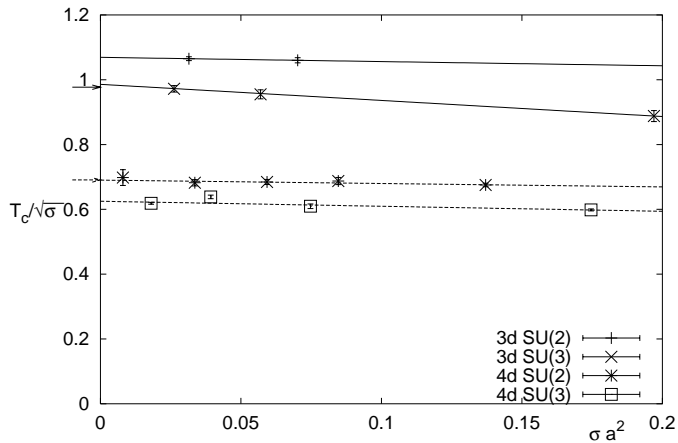, width=90mm }
  \caption{The cut-off dependence of $T_c/\sqrt{\sigma}$}
  \label{fig:TcSigma}
\end{figure}

In figure \ref{fig:TcSigma} we plot $T_c/\sqrt{\sigma}$ vs. $\sigma a^2$ for
SU(2) and SU(3) in 3 and 4 dimensions. The lines give an idea of the continuum
extrapolation. The arrows mark the values predicted from the \NGsm.

\section{Conclusions}

The critical couplings are well determined for SU(2) and SU(3) in 2+1
dimensions for $\Nt = 2, 4, 6$. 
They agree for SU(2), $\Nt=4$ and SU(3), $\Nt=2$ with the literature.

The critical exponents are in agreement with the values predicted from the
reduced model. They are definitly not the meanfield exponents. 

In SU(3) the string tension is well determined on a wide range 
of $\beta$ values and can be described by a cubic fit in $1/\beta$.
 
For SU(3) in 3 dimensions $T_c/\sqrt{\sigma}$ is in excellent agreement with
the \NGsm.

For SU(2) $T_c/\sqrt{\sigma}$ is 7\% off the predicted value.

{\bf Acknowledgement} 
We gratefully acknowledge support from the DFG 
under contract Pe 340/3-3.
Part of the calculations were performed on the DFG-Quadrics at 
Bielefeld (Pe 340/6-2).

\end{document}